\documentclass[twocolumn,showpacs,prappl,amsfonts,amsmath,amssymb,floatfix]{revtex4}
\usepackage{color}
\usepackage{hhline}
\usepackage{mathrsfs}
\usepackage{graphicx}
\usepackage{dcolumn}
\usepackage{bm}
\usepackage{multirow}
\usepackage{booktabs}
\usepackage{afterpage}
\usepackage{amsmath}

\begin{document}

\title{Two-Dimensional Valley Electrons and Excitons in Noncentrosymmetric 3R MoS$_{2}$}

\author{Ryosuke Akashi$^{1,2}$}
\altaffiliation[Present address: ]{Department of Physics, The University of Tokyo,  Tokyo 113-0033, Japan}
\author{Masayuki Ochi$^{1,2}$}
\author{S\'andor Bord\'acs$^{3,4}$}
\author{Ryuji Suzuki$^{4}$}
\author{Yoshinori Tokura$^{1,4}$}
\author{Yoshihiro Iwasa$^{1,4}$}
\author{Ryotaro Arita$^{1,2}$}

\affiliation{$^1$RIKEN Center for Emergent Matter Science (CEMS), Wako, Saitama 351-0198, Japan}
\affiliation{$^2$JST ERATO Isobe Degenerate $\pi$-Integration Project, Advanced Institute for Materials Research (AIMR), Tohoku University, Sendai, Miyagi 980-8577, Japan}
\affiliation{$^3$Department of Physics, Budapest University of Technology and Economics, 1111 Budapest, Hungary}
\affiliation{$^4$Quantum-Phase Electronics Centre (QPEC) and Department of Applied Physics, The University of Tokyo, Tokyo 113-8656,
Japan}

\date{\today}
\begin{abstract}
We find that the motion of the valley electrons -- electronic states close to the K and ${\rm K'}$
points of the Brillouin zone -- is confined into two dimensions when the layers of MoS$_{2}$ form the 3R stacking, while in
the 2H polytype the bands have dispersion in all the three dimensions. According to our first-principles band
structure calculations, the valley states have no interlayer hopping in 3R-MoS$_{2}$, which is proved to be the consequence of the rotational
symmetry of the Bloch functions. By measuring the
reflectivity spectra and analyzing an anisotropic hydrogen atom model, we confirm that the valley
excitons in 3R-MoS$_{2}$
have two-dimensional hydrogen-like spectral series, and the spreads of the wave function are smaller than the interlayer distance.
In contrast, the valley excitons in 2H-MoS$_{2}$ are well described by the three-dimensional model and thus not confined in a single layer. Our results indicate that
the dimensionality of the valley degree of freedom can be controlled simply by the stacking geometry, which can be utilized
in future valleytronics.
\end{abstract}
\pacs{73.22.-f, 71.20.Nr, 71.35.-y, 78.20.-e}

\maketitle
\section{Introduction}
\begin{figure}[t]
 \begin{center}
  \includegraphics[scale=.40]{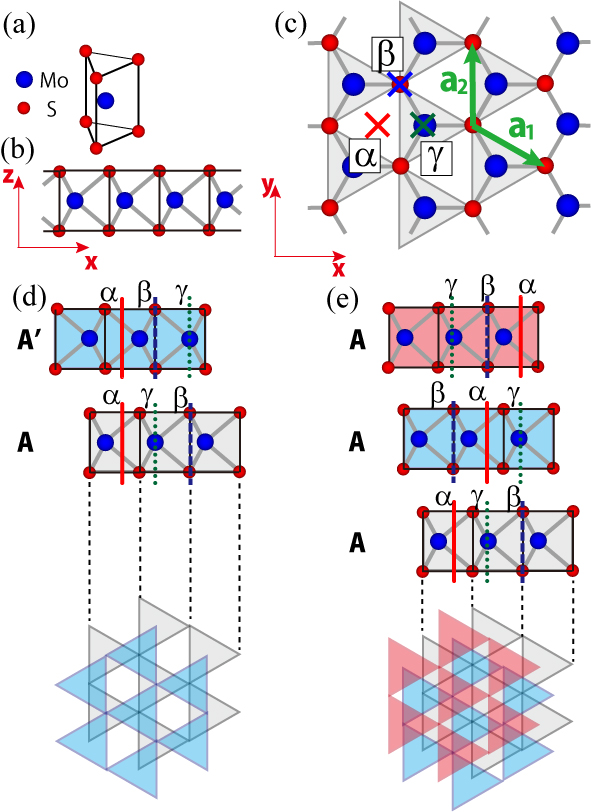}
  \caption{(a) Trigonal prismatic unit cell of monolayer MoS$_{2}$; (b)--(c) side and top views of the monolayer,
  where the trigonal
  prisms are depicted as shaded triangles. $\alpha$, $\beta$ and $\gamma$ represent inequivalent three-fold rotational axes and
  $\mathbf{a_{1}}$ and $\mathbf{a_{2}}$ are primitive lattice vectors. (d)--(e) side view (top) and top view (bottom) of the 2H and 3R stackings, respectively.
  Rotational axes for each layer are also indicated. }
  \label{fig:MoS2-struct}
 \end{center}
\end{figure}

The monolayer systems of group-VI dichalcogenides,
$MX_{2}$~($M$=Mo,~W;~$X$=S,~Se,~Te)~\cite{Novoselov-PNAS2005} have received considerable interest as unique alternatives of graphene for their various intriguing properties~\cite{Wang-Kis-review2012}: direct band gap of 2~eV~\cite{Mak-Heinz-PRL2010,Splendiani-Wang-NANOlett2010,Li-Galli-JPhysChem2007,Lebegue-Eriksson-PRB2009,Ding-Tang-PhysicaB2011,Kuc-Zibouche-Heine-PRB2011,Liu-Kumar-IEEE2011,Zhu-Schwingenscholgl-PRB2011,Ataca-Ciraci-JPhysChem2011,Kadantsev-Hawrylak-SSComm2012}, high photoluminescence yield~\cite{Mak-Heinz-PRL2010, Splendiani-Wang-NANOlett2010}, high on/off switching ratio in field
effect transistors~\cite{Kis-MoS2-single, Yoon-MoS2-single}, and electric field-induced superconductivity \cite{Taniguchi-super,
Ye-Akashi-super}. Most notably, their staggered-honeycomb-like lattice structure (the structure and the unit cell are shown in Fig.~\ref{fig:MoS2-struct} (a)--(c)) hosts a model system for valley-dependent phenomena~\cite{Xiao-PRL2012} originally proposed for graphene~\cite{Rycerz-valley-filter,Xiao-graphene-PRL2007,Yao-graphene-PRB2008}. The bottom of the conduction band and the top of the valence band are located at the K points of the hexagonal Brillouin zone in $MX_{2}$. Since the K and ${\rm K}'$$=-$K points are not equivalent, the electronic states in the opposite pockets can carry an additional quantum number, the valley index. Furthermore, the spin-orbit coupling results in sizable valley-dependent spin splitting at the valence top (VT). On the basis of the strong coupling between the spin, orbital and valley degrees of freedom, the control of the carrier population of each valley by orbital/spin-sensitive probes has been proposed. Indeed, valley-selective excitation of the electrons and excitons by circularly polarized light has been demonstrated~\cite{Mak-Heinz-NNano2012, Zeng-Cui-Nnano2012, Cao-Feng-NComm2012, Jones-Xu-WSe2-NNano2013, Mak-valley-Hall-Science2014}. Further understanding and stable control of the valley-dependent electronic properties could bring us to the {\it valleytronics}---a way of encoding information into the valley degree of freedom, which can be much faster and more efficient than conventional optoelectronics~\cite{Rycerz-valley-filter,Xiao-PRL2012,Xu-Heinz-review-NPhys2014}. 

The above progress led to renewed attention to the rich polymorphism of multilayered $MX_{2}$
~\cite{Wilson-Yoffe-review,Wang-Kis-review2012} in view of the valley physics. In inversion-symmetric bilayers with the 2H stacking
[Fig.~\ref{fig:MoS2-struct} (d)], the net valley-dependent spin polarization is absent. This property has been
proposed to be utilizable for switching of the polarization with a symmetry-breaking electric field~
\cite{Wu-efield-NPhys2013, Yuan-Saeed-NPhys2013} or mechanical bending~\cite{Ramasubramaniam-bend}. It has also been stated that
a layer degree of freedom (upper/lower layer) couples to the valley degrees of freedom~\cite{Gong-ME-NComm2013,
Jones-locking-NPhys2014} and this coupling should cause magnetoelectric effects~\cite{Gong-ME-NComm2013}. On the other hand,
very recently, Suzuki {\it et al.}~\cite{Suzuki-MoS2} observed valley-dependent spin polarization in
multilayered MoS$_{2}$ by utilizing the noncentrosymmetric 3R stacking~[Fig.~\ref{fig:MoS2-struct} (e)]. This success paves a
very different way to the control of the valley carriers: Valleytronics by engineering the stacking geometry. However, knowledge
 of the valley electronic states in the 3R systems is still scarce, which hampers further advances in the field and applications of multilayered MoS$_{2}$.

In this article, we study the valley electronic states in MoS$_{2}$ with the 3R stacking and compare them with those in
2H-MoS$_{2}$. Combining {\it ab initio} band structure calculations and group-theoretical analysis, we show that
the interlayer hopping amplitude of the valley states is exactly zero in 3R-MoS$_{2}$, i.e. the electrons are
confined within the two-dimensional (2D) layers. Furthermore, we study how this confinement affects the exciton spectrum
with an anisotropic hydrogen atom model. Finally, we compare the theoretical results to the
reflectivity spectra measured on both 3R-MoS$_2$ and 2H-MoS$_2$ compounds. The revealed mechanism of the single-layer confinement of the valley electrons respects only the crystal and orbital symmetries and therefore is commonly applicable to the family of 3R-$MX_{2}$, which should facilitate the dimensionality control of the valley states in the transition-metal dichalcogenides. 

\begin{figure}[htbp]
 \begin{center}
  \includegraphics[scale=.35]{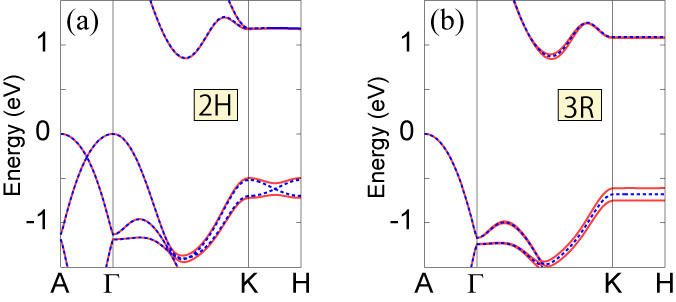}
  \caption{(a)--(b) Band structures of bulk MoS$_{2}$ with the 2H and 3R stacking, respectively, where we employed
  special points of the
  conventional hexagonal Brillouin zone (BZ) defined by $\Gamma$$=$$(0,0,0)$, ${\rm A}$$=$$(0,0,\pi/c)$, ${\rm
  K}$$=$$(0,4\pi/(3a),0)$
  and ${\rm H}$$=$$(0,4\pi/(3a),\pi/c)$ with $a$$=$$|{\bf a}_{1}|$$=$$|{\bf a}_{2}|$ and $c$ being the interlayer distance. Note that the the present BZ is twice as large as the primitive BZ of 2H-MoS$_{2}$.
  Solid (dashed) lines denote the result with (without) spin-orbit coupling. }
  \label{fig:bulk-band}
 \end{center}
\end{figure}

\section{Theory on the valley electronic states}
\subsection{First-principles calculation}
First, we calculated the band structures of bulk 2H- and 3R-MoS$_2$ using the \textsc{wien2k} code employing the full-potential linearized augmented plane-wave method~\cite{wien2k}. We used the Perdew-Burke-Ernzerhof exchange-correlation functional \cite{PBE} and included the scalar-relativistic effects~\cite{scalar-rel} and spin-orbit coupling~\cite{SO} as implemented in the \textsc{wien2k} code. Experimental lattice parameters and atomic configurations were taken from Refs.~\onlinecite{MoS2-struct-exp1} and \onlinecite{MoS2-struct-exp2}. The muffin-tin radii for Mo and
S atoms, $r_{\rm Mo}$ and $r_{\rm S}$, were set to 2.44 and 2.10\,a.u., respectively. The maximum modulus for the reciprocal
lattice vectors $K_{\rm max}$ was chosen so that $r_{\rm S}$$\times$$K_{\rm max}$~=7.00.

The calculated band structures are shown in Fig.~\ref{fig:bulk-band}. The
apparent difference seen around the $\Gamma$ point in valence bands is due to Brillouin-zone folding for the 2H case with respect to $k_{z}$$=$$\pi/(2c)$ plane and hence trivial. Notably, along the K--H path, the band at the conduction bottom (CB) is flat for the both polytypes. On the other hand, the VT bands for the 2H polytype shows small but nonzero dispersion, whereas those for 3R is flat. This feature has been first found in Ref.~\onlinecite{Suzuki-MoS2} but its origin has remained unexplored. Below, we clarify its mechanism analytically.

\subsection{Group-theoretical analysis}
In order to understand the microscopic origin of the flat bands along the K--H path, let us first look into the symmetries of the electronic
states in a monolayer of type $A$ in Fig.~\ref{fig:MoS2-struct}. The little group of K point includes three fold
rotation ($C_{3}$). Under the rotation around the axis passing through the center of the honeycomb unit [$\alpha$ in
Fig.~\ref{fig:MoS2-struct}(c)], the VT (CB) state, which is formed by Mo 4$d_{x^{2}-y^{2}}$$+$$d_{xy}$ (4$d_{z^{2}}$) orbital having
orbital angular momentum $l_{z}$$=$$-2$ ($0$), gets phase 1 ($e^{i\frac{4\pi}{3}}$)~\cite{Cao-Feng-NComm2012}.
\begin{figure}[t]
 \begin{center}
  \includegraphics[scale=.47]{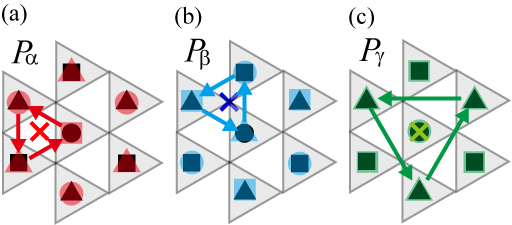}
  \caption{(a)--(c) Schematic of the phase change due to site permutation. Real-space configuration of the phase of the K-point
  Bloch state [$\exp(i{\bf K}\cdot{\bf r})$] is represented, where circle, triangle and square represent the three phase values
  $1$, $e^{i\frac{2\pi}{3}}$ and
  $e^{i\frac{4\pi}{3}}$, respectively. Large shaded triangles correspond to those in Fig~\ref{fig:MoS2-struct}. Phase
  configurations before (after) the corresponding permutation $P_{j}$ are represented by solid (shaded) symbols. Crosses
  indicate
  the rotational centers $j=\alpha$, $\beta$ and $\gamma$, whereas arrows represent how the sites permute with $P_{j}$.}
  \label{fig:rotate-scheme}
 \end{center}
\end{figure}
\begin{table}[b]
\caption[t]
{Phases $\delta^{b}_{j}$ of the eigenvalues $e^{2\pi i\delta^{b}_{j}}$ of $C_{3;j}\equiv R P_{j}$ around different rotational
axes ($j=$$\alpha$, $\beta$ and $\gamma$) for the K-point states ($b$=VT, CB).}
\begin{center}
\label{tab:phases}
\tabcolsep = 1mm
\begin{tabular}{|l |c|c|c|c|c|c||c|c|c|c|c|c|} \hline
 &\multicolumn{6}{c||}{A}&\multicolumn{6}{c|}{A'} \\ \hline
$b$ &\multicolumn{3}{c|}{VT}&\multicolumn{3}{c||}{CB} &\multicolumn{3}{c|}{VT}&\multicolumn{3}{c|}{CB} \\ \hline
 $j$&$\alpha$ & $\beta$ &  $\gamma$ &$\alpha$ & $\beta$ &  $\gamma$ &$\alpha$ & $\beta$ &  $\gamma$ &$\alpha$ & $\beta$ &
 $\gamma$  \\
  \hline
 $P_{j}$ & $\frac{2}{3}$ & $\frac{1}{3}$  & $0$ &$\frac{2}{3}$ & $\frac{1}{3}$  & $0$& $\frac{1}{3}$ &$\frac{2}{3}$ & 1
 &$\frac{1}{3}$ &$\frac{2}{3}$ & $0$  \\ \hline
 $R$ & \multicolumn{3}{c|}{$\frac{1}{3}$}& \multicolumn{3}{c||}{$0$}&\multicolumn{3}{c|}{$\frac{2}{3}$}
 &\multicolumn{3}{c|}{$0$}
 \\ \hline
 $C_{3;j}$ & $0$ &$\frac{2}{3}$  & $\frac{1}{3}$ &$\frac{2}{3}$ & $\frac{1}{3}$  & $0$&  $0$ & $\frac{1}{3}$ &$\frac{2}{3}$
 &$\frac{1}{3}$ &$\frac{2}{3}$ & $0$\\
 \hline
\end{tabular}
\end{center}
\end{table}

As a matter of fact, there are two more rotational axes inequivalent to $\alpha$: $\beta$ and $\gamma$ passing through S and Mo sites,
respectively [Fig.~\ref{fig:MoS2-struct}(c)]. In order to calculate the phase factors for these $C_{3}$ axes, let
us decompose the
$C_{3}$ rotation into two successive operations: permutation of the atomic sites $P$ and rotation of the local
coordinates at each lattice point $R$. The $C_{3}$ around $j$ axis is defined as $C_{3;j}\equiv R P_{j}$
($j=\alpha,
\beta, \gamma$), where $P_{j}$ denotes the corresponding site permutation. The contribution to the total phase from $R$ is
simply
$e^{i\theta l_{z}}$ with $\theta$=$2\pi/3$. The contribution from $P_{j}$ can be calculated by examining how the
Bloch wave function transforms under the site permutation. In Fig.~\ref{fig:rotate-scheme}, the phase of the K-point Bloch state
is
depicted, where three values ($1$, $e^{i\frac{2\pi}{3}}$ and $e^{i\frac{4\pi}{3}}$) are represented by circle, triangle and
square, respectively. By operating $P_{\alpha}$, e.g., the configuration depicted with solid symbols is transformed to that with
shaded symbols [panel (a)]. Comparing these symbols on each site, the phase contribution $e^{i\frac{4\pi}{3}}$ for
$P_{\alpha}$ is deduced. Applying similar arguments to $j$=$\beta$ and $\gamma$, and to the monolayer of type $A'$, we get
the results summarized in Table~\ref{tab:phases}. Note that the phases for $A'$ are inverse of those for $A$ because their valley
states are related by spatial inversion and time reversal.

On the basis of the axis-dependent phases derived above~\cite{footnote-Yao}, let us next consider the interlayer hopping amplitude defined by 
\begin{eqnarray}
t_{b}\equiv \langle \Psi^{b}_{{\bf K},1} | \mathcal{H}| \Psi^{b}_{{\bf K},2} \rangle \ \ (b={\rm VT},~{\rm CB})
\end{eqnarray}
 for the bulk material. Here, $|\Psi^{b}_{{\bf K},L} \rangle$ denotes the K-point
Bloch functions at $L$th layer, which are eigenstates in the limit where the layer is isolated from others. $\mathcal{H}$ is the
total Hamiltonian of the bulk crystal. Suppose $\mathcal{C}_{3}$ denotes a certain three-fold rotation under which
$\mathcal{H}$ is invariant, then
\begin{eqnarray}
\langle \Psi^{b}_{{\bf K},1} | \mathcal{H}| \Psi^{b}_{{\bf K},2} \rangle
&=&
\langle \Psi^{b}_{{\bf K},1} |\mathcal{C}^{-1}_{3} \mathcal{C}_{3} \mathcal{H}\mathcal{C}^{-1}_{3} \mathcal{C}_{3}|
\Psi^{b}_{{\bf K},2} \rangle
\nonumber \\
&=&
(\langle \Psi^{b}_{{\bf K},1} |C^{-1}_{3;j} ) \mathcal{H} (C_{3;j'}| \Psi^{b}_{{\bf K},2} \rangle)
\nonumber \\
&=&
\exp[-2\pi i \Delta^{b}_{12;jj'}] \langle \Psi^{b}_{{\bf K},1} | \mathcal{H}| \Psi^{b}_{{\bf K},2} \rangle
\label{eq:hopping}
.
\end{eqnarray}
Here, $\Delta^{b}_{12;jj'}$$=$$\delta^{b}_{1;j}-\delta^{b}_{2;j'}$ and $\delta^{b}_{L;j}$ denotes the phase summarized in
Table~\ref{tab:phases} for the $L$th layer. Note that $j$ and $j'$ in the second and last lines depend on the stacking pattern
and rotational axis of $\mathcal{C}_{3}$. In the 2H stacking, without loss of generality, we can assume the first and second layers to be
$A$ and $A'$, respectively. In this case, the possible combinations $(j, j')$ are $(\alpha, \alpha)$,~$(\beta, \gamma)$ and
$(\gamma, \beta)$ [Fig.~\ref{fig:MoS2-struct} (d)], all of which yield $\Delta^{b}_{12;jj'}$$=$$0 \ (b$$=$${\rm VT})$ and $2/3 \
(b$$=$${\rm CB})$, respectively. In the 3R stacking, where only layers of $A$ type are stacked, $(j, j')$$=$$(\alpha, \beta)$,~$(\beta,
\gamma)$,~$(\gamma, \alpha)$ [Fig.~\ref{fig:MoS2-struct} (e)], and therefore $\Delta^{b}_{12;jj'}$$=$$1/3 \ (b={\rm VT, CB})$.
According to Eq.~(\ref{eq:hopping}), the nonzero $\Delta^{b}_{12;jj'}$ forces $t_{b}$ to be zero, which consistently explains
the dispersionless bands along the K--H direction in the 3R polytype. 

\section{Valley excitons}
\subsection{Experiment}
\begin{figure}[t]
 \begin{center}
  \includegraphics[scale=.45]{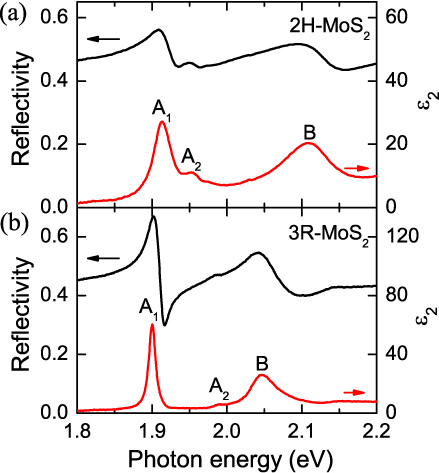}
  \caption{(a)--(b) Reflectivity spectra of bulk 2H- and 3R-MoS$_2$ measured at $T$=20\,K. The imaginary part of the
  dielectric function was obtained from the reflectivity spectra by Kramers-Kronig transformation. For the $A$ exciton branch,
  $1s$ and $2s$ excitons are labeled as $A_1$ and $A_2$, while the fine structure of the $B$ exciton cannot be resolved.}
  \label{fig:reflectivity}
 \end{center}
\end{figure}

The difference in the band structures of 2H-
and 3R-MoS$_2$ should cause different optical properties of their valley excitations. To verify this experimentally, we measured the
normal incidence reflectivity spectra on the cleaved (001) surfaces of both polytypes at $T$=20\,K (see
Fig.~\ref{fig:reflectivity}). The details of the crystal growth were reported in Ref.~\onlinecite{Suzuki-MoS2}. Since the
reflectivity were measured in the limited photon energy range of $E$=1.5--3\,eV, the spectra were extended up to 30\,eV using the
data published for 2H-MoS$_2$~\cite{MoS2_KK} to facilitate proper Kramers-Kronig transformation. 

The imaginary part of the
dielectric function calculated from the reflectivity is shown in Fig.~\ref{fig:reflectivity}. Although the two main features in the $\varepsilon_2$ spectra, the $A$ and $B$ exciton peaks corresponding to the optical excitations from the spin-orbit split VT to the CB, look similar, there are several important differences between the two spectra. First, the energy separation between the $A_1$ and the $B$ peaks is larger for the 2H compound (197\,meV) compared with the 3R polytype (148\,meV), which agrees with the theoretical prediction for the splitting at the VT (see Fig.~\ref{fig:bulk-band} and Ref.~\onlinecite{Suzuki-MoS2}). Beside the main excitonic resonances, smaller peaks were observed on the high energy shoulder of the $A$ exciton, labeled as $A_2$. In a hydrogen-atom-like model of excitons, $A_1$ and $A_2$ are assigned to the 1$s$ and 2$s$ bound states of the electron-hole pair. In the 3R case, $A_1$ and $A_2$ peaks are more separated and their intensity ratio is larger than in the 2H structure.
\begin{table}[b]
\caption[t]
{Parameters adopted for Eq.~(\ref{eq:exciton-model}) extracted from \textit{ab initio} band structure calculations: dielectric
constants and reduced effective masses of excitons at the K-point (in units of electron mass $m_0$).}
\begin{center}
\label{tab:parameters-in-the-model}
\footnotesize{\tabcolsep = 1mm
\begin{tabular}{lcccccc} \hline\hline
  &  & $\varepsilon_{\perp}$& $\varepsilon_{\parallel}$ & & $m^{\ast}_{\perp}$ & $m^{\ast}_{\parallel}$ \\
  \hline
 2H & Ours & 16.0 & 10.2 & & 0.27 & 1.87\\
       & Ref.~\onlinecite{epsilon-ref1} & 13.5 & 8.5 &  & - & - \\
      & Ref.~\onlinecite{epsilon-ref2} & 16 & 10 &  & - & - \\
       & Ref.~\onlinecite{mass-ref1} & - & - &  & 0.22 (K--$\Gamma$), 0.23 (K--M) & 1.7 \\
 3R & Ours & 15.9 & 10.1 & & 0.24 & 28 \\
 \hline\hline
\end{tabular}
}
\end{center}
\end{table}

\subsection{Model analysis}
To understand these differences, we employed an anisotropic hydrogen atom model \cite{anisotropic-hydrogen}:
\begin{eqnarray}
&&
\left[
\!-
\frac{\nabla^{2}_{x}\!\!+\!\!\nabla^{2}_{y}}{2m^{\ast}_{\perp}}
\!-\!
\frac{\nabla^{2}_{z}}{2m^{\ast}_{\parallel}}
\!-\!
\frac
{1}
{\sqrt{\varepsilon_{\perp}\varepsilon_{\parallel}(\!x^{2}\!+\!y^{2})\!\!+\!\!\varepsilon_{\perp}^{2}z^{2}}}
\right]
\varphi({\bf r})
\nonumber \\
&&\hspace{150pt}
=E\varphi({\bf r}),
\label{eq:exciton-model}
\end{eqnarray}
where $\varepsilon_{\perp}$ and $\varepsilon_{\parallel}$ are the components of the dielectric constant tensor perpendicular and
parallel to the $z$-direction respectively, and $m^{\ast}_{\perp}$ and $m^{\ast}_{\parallel}$ are those of the reduced
effective-mass tensor of excitons at the K-point. Note that in a monolayer system, the screened electron-hole interaction is
rather different from that in Eq.~(\ref{eq:exciton-model}) owing to the surrounding vacuum~\cite{Keldysh,nonRydberg}. Bulk
environment allows us to use the present model as in early studies~\cite{MoS2-thin-crystals,exciton-spectr}. 

The values of 
$\varepsilon_{\perp}$, $\varepsilon_{\parallel}$, $m^{\ast}_{\perp}$ and $m^{\ast}_{\parallel}$ were extracted from {\it ab
initio} calculations using the {\sc wien2k} code. Dielectric constants $\varepsilon$ were calculated using the random-phase approximation~\cite{Ehrenreich,Adler,Wiser},
\begin{eqnarray}
\varepsilon_{\alpha}
\!=\!
1
\!+\!
\frac{8\pi}{V}
\lim_{q\rightarrow 0}
\!\!
\frac{1}{q^{2}}
\sum_{{\bf k}\sigma}
\!\sum_{i}^{\rm occ.}
\!\sum_{j}^{\rm unocc.}
\!\!
\frac
{|\langle \phi_{j{\bf k}+q{\bf e}_{\alpha}\sigma}|e^{iq{\bf e}_{\alpha}\cdot {\bf r}}|\phi_{i{\bf k}\sigma}\rangle|^{2}}
{\varepsilon_{j{\bf k}+q{\bf e}_{\alpha}\sigma}-\varepsilon_{i{\bf k}\sigma}},
\nonumber \\
\end{eqnarray}
where $V$ denotes the volume of the simulation cell, $i$ and $j$ are indices of the occupied and unoccupied bands, respectively, ${\bf k}$ is the wave vector, ${\bf e}_{\alpha}\ (\alpha=x,y,z)$ denotes the unit vectors, $\sigma$ denotes the spin index, $\phi_{i{\bf k}\sigma}$ denotes the Kohn-Sham wave functions, and $\varepsilon_{i{\bf k}\sigma}$ denotes their orbital energies.
The effective masses $m^{\ast}$ were estimated from a parabolic curve fitting of the band dispersions of electrons and holes around the K point within the range of $\sim 0.02\ \AA^{-1}$ ($\sim 0.03\ \AA^{-1}$ for $m^{\ast}_{\parallel}$ of the 3R polytype). To obtain the effective masses accurately, $r_{\rm
Mo}$, $r_{\rm S}$, and $r_{\rm S}$$\times$$K_{\rm max}$ were set to 2.20\,a.u., 1.89\,a.u., and 9.00, respectively.

Calculated dielectric constants and reduced effective masses are presented in Table~\ref{tab:parameters-in-the-model}, which agree well with previous calculations~\cite{epsilon-ref1, epsilon-ref2, mass-ref1}. Since both VT and CB are flat along the K--H line, $m^{\ast}_{\parallel}$ is quite large in the 3R stacking, whereas it is small
in the 2H stacking in which only CB is flat.
Note that the effective masses were determined in the close vicinity of the K point, 
and the valence and conduction bands in 3R-MoS$_2$ have less than 2\,meV dispersion along the K--H line. Since
this range is much smaller than the binding energy of $1s$ or $2s$ exciton, in the following analyses, we regard the 3R bands
completely flat, i.e. $m^*_{\parallel}$$=$$\infty$.

\begin{table}[b]
\caption[t]
{Exciton binding energies of $1s$ exciton $E_{1s}$, $2s$ exciton $E_{2s}$, their difference $\Delta E=E_{2s}-E_{1s}$ (all in
meV), and a relative ratio of oscillator strengths of $1s$ and $2s$ excitons $f_{1s}/f_{2s}$. Theoretical values are obtained
from the anisotropic hydrogen atom model (Eq.~(\ref{eq:exciton-model})). Experimental values are obtained from a Lorentz model
fit for the reflectivity spectra (Fig.~\ref{fig:reflectivity}). The detail of the model fitting is shown in Appendix~\ref{sec:fitting}.}
\begin{center}
\label{tab:exciton-spectr-calc}
\footnotesize{\tabcolsep = 1mm
\begin{tabular}{cccccc} \hline\hline
  & & $-E_{1s}$ & $-E_{2s}$ &$\Delta E(=E_{2s}-E_{1s})$ & $f_{1s}/f_{2s}$ \\
  \hline
 2H & Theory & 34 & 8.9 & 25 & 8.2 \\
      & Expt. & & & 41.7$\pm$0.8 & 3.7$\pm$0.5 \\
 3R & Theory & 81 & 8.9 & 72 & 27 \\
      & Expt. & & & 89.6$\pm$0.3 & 39$\pm$5 \\
 \hline\hline
\end{tabular}
}
\end{center}
\end{table}

Within the model of Eq.~(\ref{eq:exciton-model}), the binding energies and oscillator strengths of the $1s$ and $2s$
excitons were estimated. For the 2H case, we referred to a previous report presenting a numerical solution of the
model for a range of parameters~\cite{exciton-spectr}. For the 3R case, we instead took the exact solution of the 2D hydrogen
atom according to the above consideration.
The ratio of the oscillator strengths was derived with the Fermi golden rule~\cite{Intensity} using the present energy
eigenvalues and the reduced effective masses. The obtained values are presented in Table~\ref{tab:exciton-spectr-calc} together
with our experimental ones extracted from fits with Lorentzian peaks (Fig.~\ref{fig:reflectivity}).
The substantial differences in the energy levels and in the oscillator strengths between the exciton
spectra of the 2H and 3R structures were captured by the model calculation. Since all the parameters in
the 2H and 3R cases are similar to each other except $m^{\ast}_{\parallel}$, we attribute
the differences solely to the 2D confinement of the electrons and holes in the 3R compound.

To see another aspect of the 2H and 3R excitons with respect to their dimensionality, we examined the exciton radii by using a
simple trial wave function \cite{exciton-variational} for the $1s$ eigenstate of Eq.~(\ref{eq:exciton-model}),
\begin{equation}
\varphi_{1s}(\mathbf{r})=\frac{1}{\sqrt{\pi r_{\perp}^2 r_{\parallel}}} \mathrm{exp}
\left[-\sqrt{(x^2+y^2)/r_{\perp}^2+z^2/r_{\parallel}^2} \right] ,
\end{equation}
with the parameters $r_{\perp}$ and $r_{\parallel}$, which represent exciton in-plane and out-of-plane radii, respectively,
optimized in a variational manner. 
For the 3R exciton, we used a finite $m^*_{\parallel}$ listed in Table~\ref{tab:parameters-in-the-model} to estimate an upper
bound of $r_{\parallel}$.
The obtained values of $r_{\perp}$ are almost the same values for the 2H and 3R structures: 19 and 15\,\AA, respectively.
In contrast, the values of $r_{\parallel}$ are 8.9 and 2.7\,\AA\ for the 2H and 3R structures, respectively.
Comparing these values with the interlayer distance (6.1\,\AA\ for both structures), we can see that the 3R exciton is tightly
confined in a single layer while the 2H exciton extends to the neighboring layers.

\section{Discussions}
The present results establish the 2D nature of the valley exciton in bulk MoS$_{2}$ with the 3R stacking. Note that an
apparently similar confinement effect has also been discussed for the 2H stacking as the spin-layer locking
effect~\cite{Gong-ME-NComm2013, Jones-locking-NPhys2014}. Since the magnitude of confinement is governed by the factor $\lambda/\sqrt{\lambda^{2}+t^{2}_{b}}$ with $\lambda$ being the spin-orbit coupling coefficient~\cite{Gong-ME-NComm2013}, its strength essentially relies on atomic number of $M$. Distinctively, the layer-confinement effect in the 3R stacking is a quantum interference effect caused by the symmetry of the crystal and therefore is guaranteed to be relevant in general $MX_{2}$ systems with the same structure. The complete confinement of the 3R valley states should also provide us with an arena of exploring novel interplay of low dimensionality and valley-physics.

Compared to the 2H or monolayer MoS$_{2}$, the 3R polytype has a significant merit that valley polarized carriers are more robust in the latter form. The carriers are protected from relaxation processes resulting from interlayer transitions, contrary to the 2H systems. In the previous work~\cite{Suzuki-MoS2}, we have found an indication that interlayer transitions are much suppressed and the valley excitons have much longer lifetime in 3R-MoS$_{2}$. Now we have clarified that the interlayer hopping amplitude is {\it exactly} zero and thus the protection is robust; also, it is common to all 3R-$MX_{2}$. Note that the 2D states in the 3R systems are different from those in the monolayer systems. In the former, the thickness of the sample allows us to generate valley polarized carriers away from the substrate, which will reduce impacts of the substrate potential. Hence, for example, we can use the 3R crystals as a medium (or lead wires, e.g.) for the valley device, through which the valley-polarized carriers are robustly transmitted. The unique property of the 3R stacking could thus bring about more effective valley-information transmission and generation toward dissipation-less low-energy optoelectronic devices. 

\section{Summary}
To summarize, we showed rigorous absence of the interlayer hopping for the valley electrons in 3R-MoS$_{2}$, which
was substantiated with a group theoretical argument. Through the measurement of the reflectivity spectra and analysis with an
anisotropic hydrogen atom model, the absence of the hopping has been shown to make the exciton absorption spectrum 2D hydrogen-like and the exciton wave function confined in a single layer. Our results exemplify the control of spatial dimensionality
of the valley excitations in 2H- and 3R-$MX_{2}$, which may help to develop new valleytronics devices.

\begin{acknowledgments}
This work was supported by the Strategic International Collaborative Research Program (SICORP-LEMSUPER), Japan Science and Technology Agency, Funding Program for World-Leading Innovative R~\&~D on Science and Technology (FIRST Program) and Grant-in-Aid for Scientific for Specially Promoted Research (No. 25000003) from JSPS. R. S. was supported by Leading Graduate Program (MERIT) from MEXT. S. B. was supported by Hungarian Research Funds OTKA PD111756 and K108918.
\end{acknowledgments}

\begin{appendix}
\section{FITTING OF THE IMAGINARY PART OF THE DIELECTRIC FUNCTION}
\label{sec:fitting}
\begin{figure*}[t]
 \begin{center}
  \includegraphics[scale=0.7]{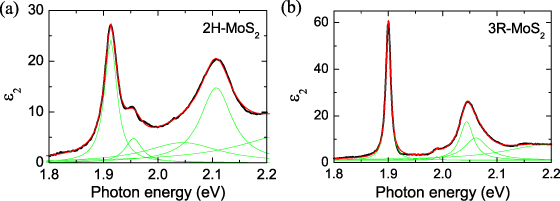}
  \caption{(a)-(b) Fit of the imaginary part of the dielectric function in 2H- and 3R-MoS$_2$, respectively. The dielectric function deduced from the reflectivity (black lines) can be fitted by a sum of Lorentzian peaks (red lines). The green lines show the contribution of the respective peaks.}
  \label{fig:SuppFit}
 \end{center}
\end{figure*}
The imaginary part of the dielectric function deduced from the measured reflectivity is fitted by a sum of Lorentzian peaks. The fits shown in Fig.~\ref{fig:SuppFit} describe the experimental spectra well. The contributions from the higher energy transitions are also included.
\end{appendix}


\begin{thebibliography}{999}
\bibitem{footnote-Yao} Note that there are also recent publications~\cite{Liu-Yao-qdot,Liu-review} utilizing the
    difference of the phase with $C_{3;\beta}$ and $C_{3; \gamma}$ (according to a different notation). The former have addressed coupling between the K and ${\rm K'}$ points in the monolayer quantum dot. The latter have discussed the interlayer hopping for the 2H stacking, giving the result consistent with ours.
\end{thebibliography}
\end{document}